\documentclass[a4paper,11pt]{article}
\usepackage{pos}
\usepackage{bm}
\usepackage{physics}

\newcommand{\eqrefeq}[1]{Eq.~\eqref{#1}}
\newcommand{\figref}[1]{Fig.~\ref{#1}}

\title{The isoscalar non-singlet axial form factor of the nucleon from lattice QCD}

\affiliation[a]{PRISMA$^+$ Cluster of Excellence \& Institut f\"ur Kernphysik,
 Johannes Gutenberg-Universit\"at  Mainz,  D-55099 Mainz, Germany}
\affiliation[b]{Helmholtz-Institut Mainz, Johannes Gutenberg-Universit\"at Mainz,
D-55099 Mainz, Germany}
\affiliation[c]{GSI Helmholtzzentrum für Schwerionenforschung, Darmstadt (Germany)}

\emailAdd{abarone@uni-mainz.de}

\author*[a]{Alessandro Barone}

\author[b,c]{Dalibor~Djukanovic} 

\author[a]{Georg~von~Hippel}

\author[a]{Jonna~Koponen}

\author[a, b]{Harvey~B.~Meyer} 

\author[a]{Konstantin~Ottnad}

\author[a,b]{Hartmut~Wittig}

\abstract{
We present our progress on the computation of the axial form factor of the nucleon with flavour structure $u+d-2s$ from lattice QCD.
We employ a set of $N_f=2+1$ CLS ensembles with $O(a)$-improved Wilson fermions and the L\"uscher-Weisz gauge action, with lattice spacings ranging from
$0.05\,\text{fm}$ to $0.086\,\text{fm}$ and pion masses spanning between $130\,\text{MeV}$ and $350\,\text{MeV}$.
We employ multiple source-sink separations and use the summation method to suppress the contamination from excited states.
We use a $z$-expansion on each ensemble to parametrize the $Q^2$-dependence of the form factor and simultaneously fit the available source-sink separations for all
$Q^2\leq 0.7 \,{\rm GeV}^2$.
We outline our analysis of the stability of the fits varying the ans\"atze and different estimations of the covariance matrix and report on our strategy for a comprehensive determination of the physical form factor.
}

\FullConference{The 41st International Symposium on Lattice Field Theory (LATTICE2024)\\
 28 July - 3 August 2024\\
Liverpool, UK\\}

\begin{document}
\maketitle

\section{Introduction}

The axial form factors $G_A(Q^2)$ of the nucleon plays a key role in understanding its \mbox{(quasi-)} elastic interactions with neutrinos. In particular,
the isoscalar channel is sensitive to elastic scattering mediated by
a $Z$, namely the strange axial form factor $G^{s}_A(Q^2)$,
whereas the isovector channel $G^{u-d}_A(Q^2)$ is sensitive to the $W$ boson exchange.
The strange form factor can be obtained combining the isoscalar singlet $G^{u+d+s}_{A}(Q^2)$ and isoscalar octect
$G^{u+d-2s}_{A}(Q^2)$ contributions. Furthermore, the strange form factor provides information about the nucleon spin, which can be decomposed into contributions from
the intrinsic quark spin, which is related to the axial charge $g_{A}$, the quark angular momentum and the gluon angular momentum~\cite{Ji:1996ek,COMPASS:2006mhr}. 

While the isovector contribution has recently received much attention from the community (see~\cite{Meyer:2016oeg,Meyer:2022mix} for a review), the isoscalar counterpart has not yet been adressed extensively~\cite{Alexandrou:2021wzv}. However, a theoretical
input for $G^s_A(Q^2)$ is becoming timely, since experiments such as MicroBooNE~\cite{Miceli:2014hva, PhysRevC.100.034604} are aiming to extract the strange form factor in the range $Q^2[\text{GeV}^2]\in[0.08, 1]$.
In this work, we focus on the computation of the axial form factor in the non-singlet channel for an extended range of $Q^2$. We report preliminary results
for the connected data $u+d$ to illustrate the complete procedure and give a preview of the full $u+d-2s$ case on a few ensembles.

\section{Lattice setup}

The form factor appears in the parametrisation of the nucleon-nucleon matrix element with the isoscalar current insertion
$A_{\mu}^{u+d-2s} = \bar{u}\gamma_\mu\gamma_5 u + \bar{d}\gamma_\mu\gamma_5 d -2\bar{s}\gamma_\mu\gamma_5 s$,  as
\begin{align}
 \bra{N(p', s')}A_{\mu}^{u+d-2s} \ket{N(p, s)} = \bar{U}^{s'}(p') \left[ \gamma_\mu\gamma_5 G_A(Q^2) - \frac{Q_\mu}{2M_N}\gamma_5 G_P(Q^2)  \right] U^s(p) \, , 
\end{align}
where $U^s(p)$ is an isodoublet Dirac spinor with momentum $p$ and spin $s$.

To address $G_A$, we calculate nucleon two- and three-point correlation functions starting from the nucleon interpolating operator
$\Psi^{\alpha}(x)=\epsilon_{abc} \left( \tilde{u}_a(x)C\gamma_5 \tilde{d}_b(x)  \right) \tilde{u}^{\alpha}_c(x)$, with $\tilde{u}(x)$ and $\tilde{d}(x)$ being the smeared up and down quark fields, respectively.
In particular, for the three-point correlators
we distinguish connected and disconnected contributions as
\begin{align}
C_{3{\rm pt},i}(\bm{q}, t,  t_s) & = 
 \sum_{\bm{x}, \bm{y}} e^{i\bm{q}\cdot\bm{y}}\Gamma_{\beta\alpha}
 \Big\langle \bar{\Psi}^{\alpha}(\bm{x}, t_s) \, A_{i}^{u+d-2s}(\bm{y}, t) \,\Psi^{\beta}(0) \Big\rangle \\ \notag
 &=  C^{\rm conn}_{3{\rm pt},i}(\bm{q}, t,  t_s) + C^{\rm disc}_{3{\rm pt},i}(\bm{q}, t,  t_s)
\end{align}
with
\begin{align}
C^{\rm disc}_{3{\rm pt},i}(\bm{q}, t,  t_s) &= \Big\langle L_i(\bm{q}, t) C_2(\bm{p}', t_s) \Big\rangle \, , \qquad
 L_i(\bm{q}, t) = - \sum_{\bm{z}} e^{i\bm{q}\cdot \bm{z}} {\rm Tr}\left[ D_q^{-1}(z,z)\gamma_i\gamma_5 \right] \, ,
\end{align}
where the connected part contains only $u+d$, and the strange quark $s$ appears only in disconnected loops.
We choose $\bm{p}' = \bm{0}$, $\bm{q} =\bm{p}'- \bm{p}=- \bm{p}$, i.e. rest frame of the final state nucleon.
We employ smeared quark fields and APE-smeared gauge fields in constructing $\Psi^{\alpha}$.
For the multiplicative renormalisation factor of the non-singlet current we refer to~\cite{Bhattacharya:2005rb}, and we take the factors
$Z_A$ from~\cite{DallaBrida:2018tpn} and $b_A$ from~\cite{Korcyl:2016ugy}, neglecting
the coefficient $\tilde{b}_A$ and $f_A$ (in the notation of~\cite{Korcyl:2016ugy}), which are assumed to be small since they parametrise
sea-quark effects.

The axial form factor $G_A$ is isolated considering the transverse component
\begin{align}
C^{T}_{3{\rm pt}, i}(\bm{q}, t,  t_s) = \epsilon^{ijk} q_j\, C_{3{\rm pt},k}(\bm{q}, t,  t_s) \propto (\bm{q}\times\bm{\gamma})_i \gamma_5 G_{A}(Q^2) \, ,
\end{align}
which is then projected into
\begin{align}
 C_{3\rm{pt}}(\bm{q}, t,  t_s) = \sum_i \frac{(\bm{q}\times\bm{s})_i}{|\bm{q}\times\bm{s}|^2} \, C^{T}_{3{\rm pt}, i}(\bm{q}, t,  t_s)  \,,\quad 
 \bm{s} = \bm{e}_3 \,, \quad
 \Gamma = \frac{1}{2}(1+\gamma_0)(1+i\gamma_5 \gamma_3) \, .
\end{align}
The signal is improved considering only momenta $|q_3|\leq \min\left( |q_1|, |q_2| \right)$, after which we can build the ratio
\begin{align}
R(\bm{q}, t, t_s) = \frac{C_{3\rm{pt}}(\bm{q}, t,  t_s) }{C_{\rm 2pt}(\bm{0}, t_s)}
  \sqrt{\frac{C_{\rm 2pt}(\bm{q}, t_s-t) C_{\rm 2pt}(\bm{0}, t) C_{\rm 2pt}(\bm{0}, t_s) }{C_{\rm 2pt}(\bm{0}, t_s-t) C_{\rm 2pt}(\bm{q}, t)C_{\rm 2pt}(\bm{q}, t_s)}} \, ,
\end{align}
which is directly related to the effective form factor $G^{\rm eff}_{A}(Q^2)$ in the limit $t_s-t\gg 0$.
 
The calculations are 
performed employing a set of $N_f=2+1$ CLS ensembles~\cite{Bruno:2014jqa}
with $O(a)$-improved Wilson fermions~\cite{Sheikholeslami:1985ij,Bulava:2013cta}
and the L\"uscher-Weisz gauge action~\cite{Luscher:1984xn},
with lattice spacings ranging from
$0.05\,\text{fm}$ to $0.086\,\text{fm}$ and pion masses ranging from $130\,\text{MeV}$ to $350\,\text{MeV}$. We refer to~\cite{Djukanovic:2022wru}
for the full details (see in particular Tab. I).

\section{Analysis strategy}

\begin{figure}[t]
 \centering
 \includegraphics[scale=0.30]{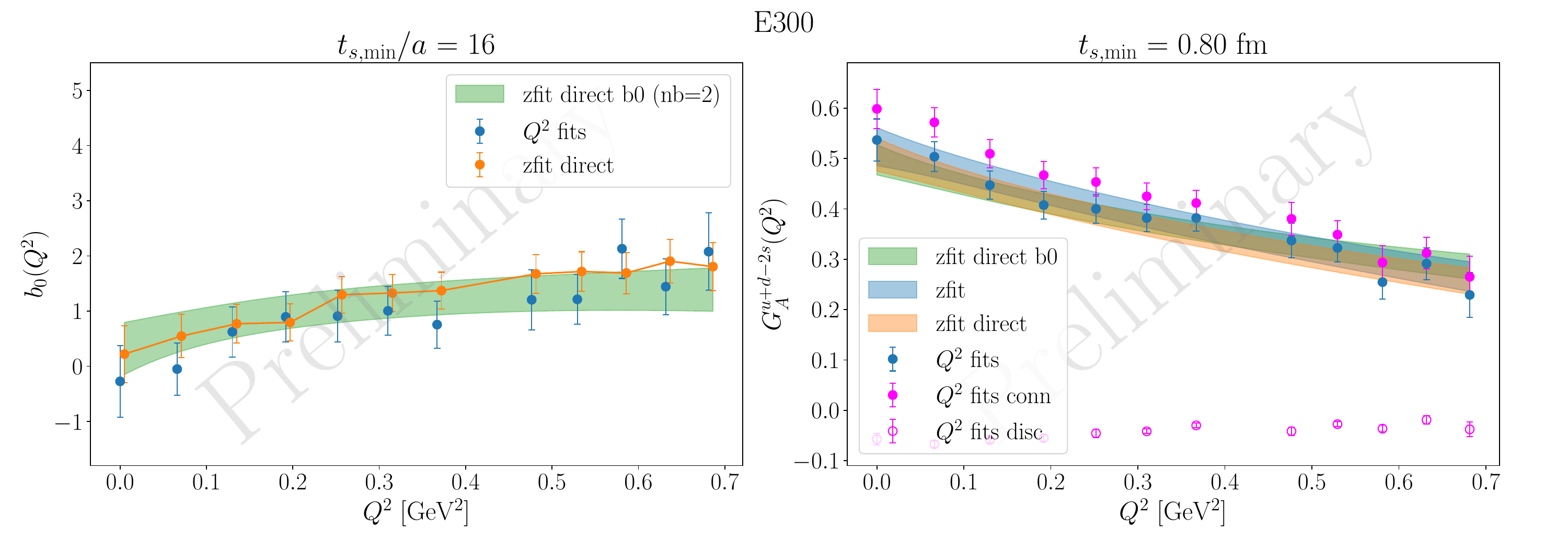}
 \caption{Comparison of different $z$-fit procedures for the ansatz in~\eqrefeq{eq:summation} for $u+d-2s$ data on E300. The blue points refer to the two-step procedure, and the orange points to the ``direct'' approach. The green bands complement the latter considering the case where $b_0$ is also parametrised by a $z$-expansion at order $n_b=2$.
 The magenta points show the contribution of the connected and disconnected data in the two-step procedure for illustration purposes.}
 \label{fig:zfit-comparison}
\end{figure}

The analysis strategy follows and extends the one presented in~\cite{Djukanovic:2022wru}. We employ the summation method~\cite{Maiani:1987by,Capitani:2012gj}
\begin{align}
\label{eq:summation}
S(\bm{q}, t_s) = a \sqrt{\frac{2E_q}{M_N+E_q}} \sum_{t=a}^{t_s-a} R(\bm{q}, t, t_s) \stackrel{t_s\gg 1}{=} b_0(Q^2) + t_s G_A(Q^2) 
 \, +\mathcal{O}(t_s e^{-\Delta t_s}) \, ,
\end{align}
which allows us to extract the form factor $G_A(Q^2)$ through a linear fit of the above expression in the source-sink separation $t_s$.
We parametrise the form factor using the $z$-expansion at order $n=2$ as
\begin{align}
\label{eq:GA-zexp}
G_A(Q^2) = \sum_{k=0}^{n} a_k z^{k}(Q^2) \, , \quad z(Q^2) = \frac{\sqrt{t_{\rm cut}+Q^2}-\sqrt{t_{\rm cut}}}{\sqrt{t_{\rm cut}+Q^2}+\sqrt{t_{\rm cut}}} \, ,
\end{align}
where we set $t_{\rm cut} = (4M_{\pi})^2$ for all the ensembles and $Q^2_{\rm max}=0.7\, \text{GeV}^2$.
The typical two-step procedure consists in extracting $G_A(Q^2)$ from a linear fit to~\eqrefeq{eq:summation} for all $t_s$, and then performing a $z$-fit over the selected points on the $Q^2$ range to extract $a_i$; instead, here we perform a single z-fit
on all data including all  $t_s \in \{t_{s,\rm min}, ...\}$ and $Q^2 \in \{0, ..., Q^2_{\rm max}\}$ to extract directly the $z$-expansion coefficients on each ensemble.
This allows us to smoothen the analysis strategy, providing a solid estimate of $a_i$ for each $t_{s, \rm min}$ with a single fit.

We compare these procedures in~\figref{fig:zfit-comparison}, considering both  the case where $b_0(Q^2)$ is treated as a fit parameter for each $Q^2$ (orange and blue points) or the case
where it is also parametrised with a $z$-expansion $b_0(Q^2) = \sum_{k=0}^{n_b} d_k z^{k}(Q^2)$ at order $n_b=2$. The plot shows that all approaches are compatible within errors.

\begin{figure}[t]
 \centering
 \includegraphics[scale=0.33]{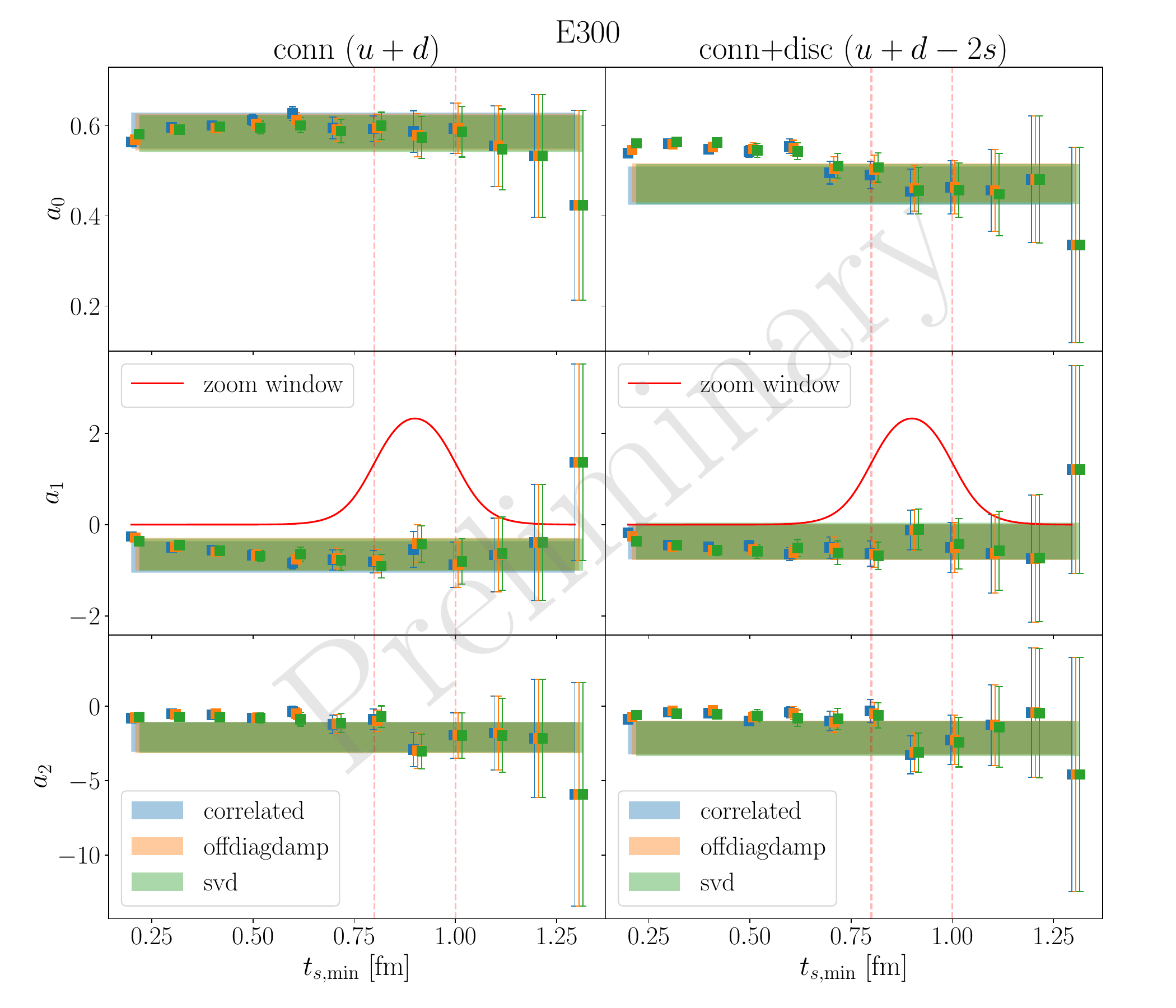}
 \caption{Window average on the coefficients $a_0, a_1, a_2$ (rows) of the $z$-expansion in~\eqrefeq{eq:GA-zexp} as a function of the minimum source-sink separation $t_{s, \rm min}$ on the ensemble E300 for the connected case (left) and the full octet case (right). The different colours refer to different approaches to regularise the covariance matrix, and the red curve is a zoom on the window function in~\eqrefeq{eq:window}. The vertical lines correspond to the choices of $t^{\rm low}_{w}$ and $t^{\rm up}_{w}$ and the horizontal bands indicate the final results of the window average on all points.}
 \label{fig:window}
\end{figure}

The fits are performed starting from a minimum value of the source-sink separation $t_{s, \rm min}$, such that the coefficients $a_i$ of the $z$-expansion depend on this choice. To obtain the final coefficients we perform a weighted average over all these values assigning the weights according to the window function
\begin{align}
\label{eq:window}
 W=\frac{1}{N_w}\left[ \tanh\left(\frac{t_s^{\rm min}-t_w^{\rm low}}{\Delta t_w} \right) - \tanh\left(\frac{t_s^{\rm min}-t_w^{\rm up}}{\Delta t_w} \right) \right] \, , \qquad
\end{align}
where $N_w$ is a normalisation factor and $t_s^{\rm low} = 0.8 \, \text{fm}$, $t_s^{\rm up} = 1\, \text{fm}$, $\Delta t_w = 0.08\, \text{fm}$ on each ensemble, in order to reduce the human bias in the procedure.

While the direct $z$-fit provides a simple solution to fitting simultaneously a large amount of data, it comes with the downside of dealing with a sizeable $N\times N$ covariance matrix, 
with $N=N_{Q^2}\times N_{t_s}$ being respectively the number of $Q^2$ and source-sink separations $t_s$ entering the fit. We therefore explore two different ways of regulating such a matrix. The first one consists in introducing a small damping $\alpha \in [0.985, 1]$ on the off-diagonal elements~\cite{Djukanovic:2022wru}; the second relies on an svd cut
to decrease the condition number of the matrix.
We compare the methods in~\figref{fig:window} against the unregulated matrix (``correlated'') to demonstrate
that the estimation of the covariance is solid, as different regularisations provide negligible differences. We quote our final results using the svd approach.

\section{Preliminary results}

\begin{figure}[t]
\hspace{-0.5cm}
\hbox{
\includegraphics[scale=0.2]{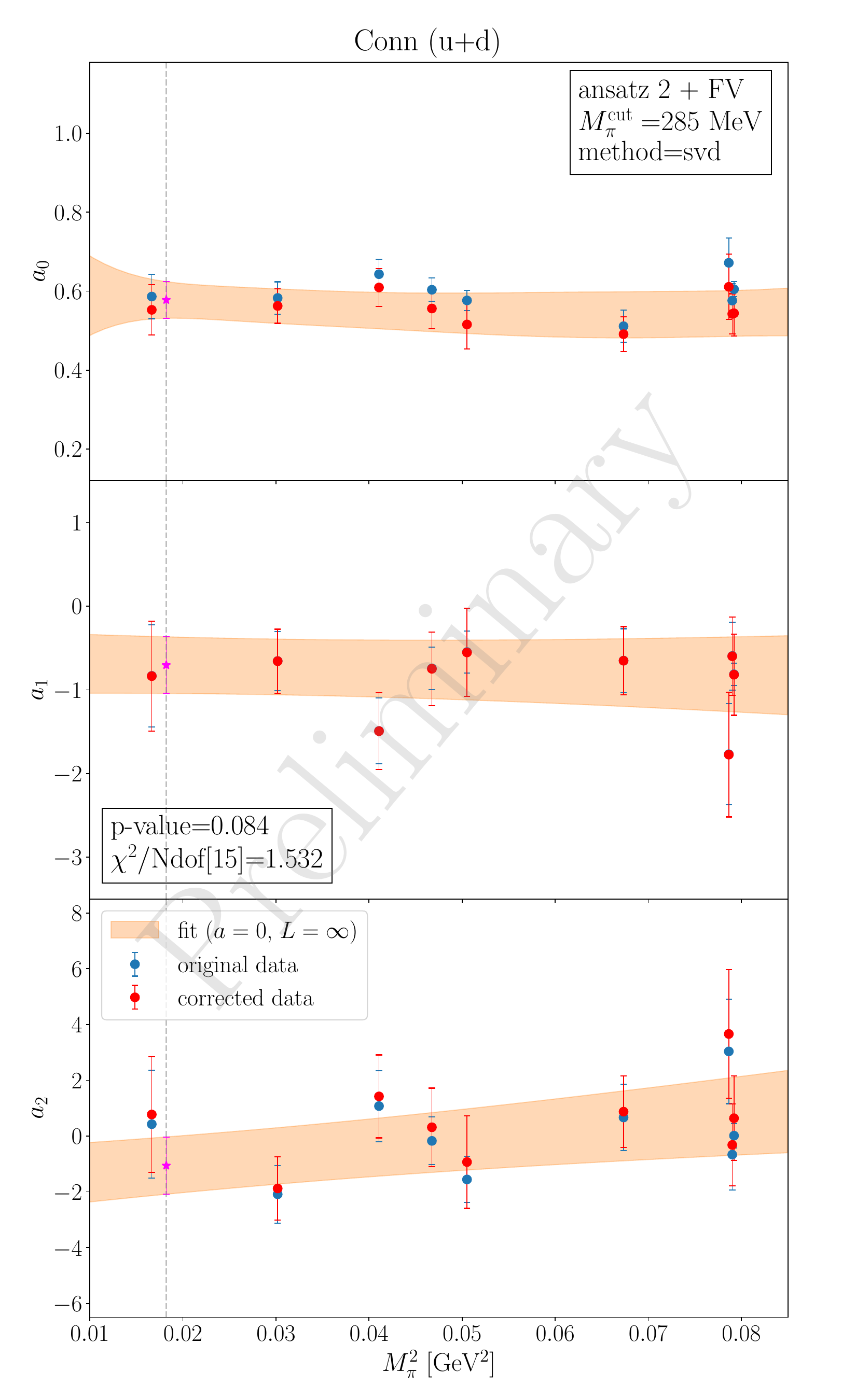}
\hspace{-0.65cm}
\includegraphics[scale=0.2]{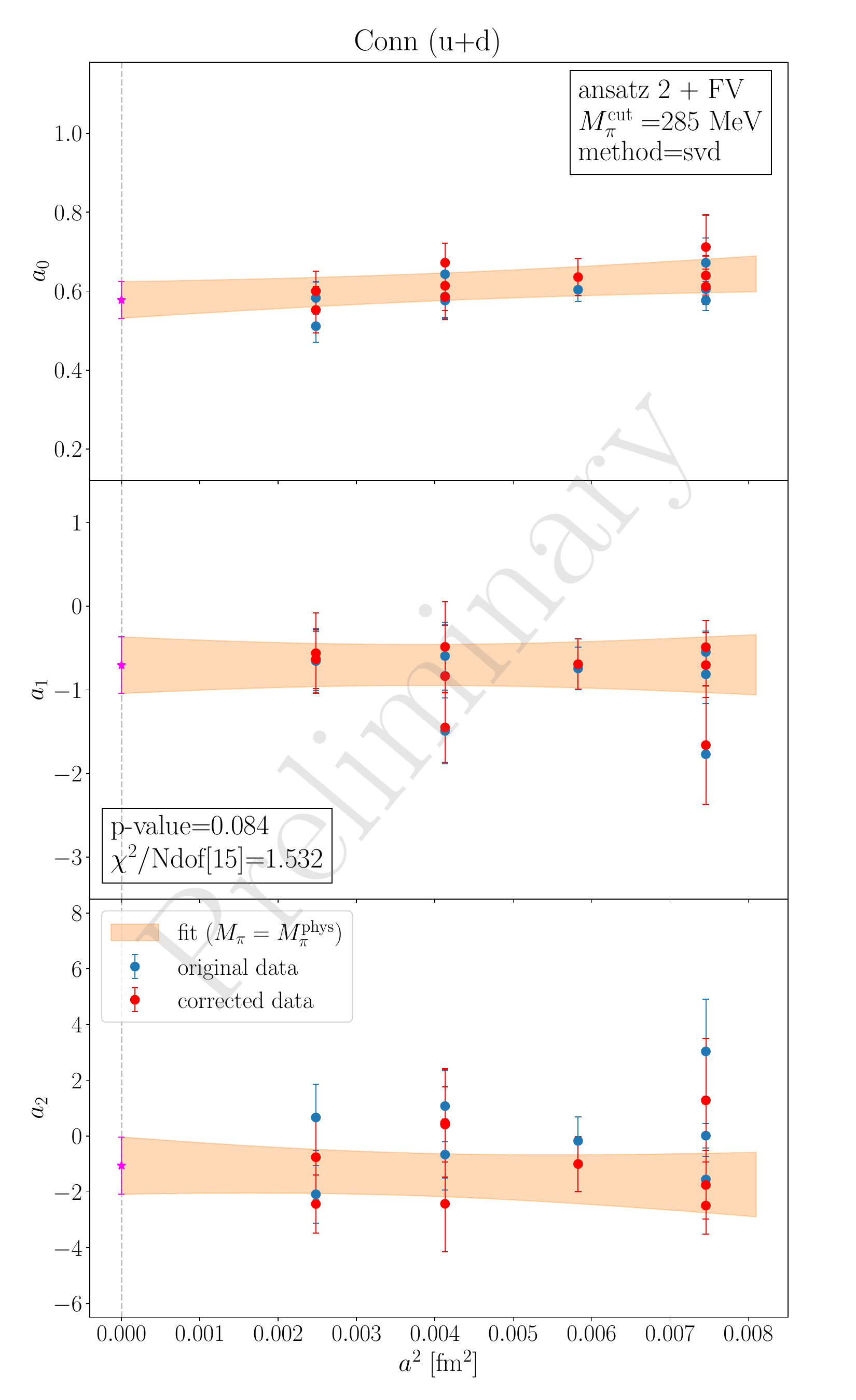}
\hspace{-0.65cm}
\includegraphics[scale=0.2]{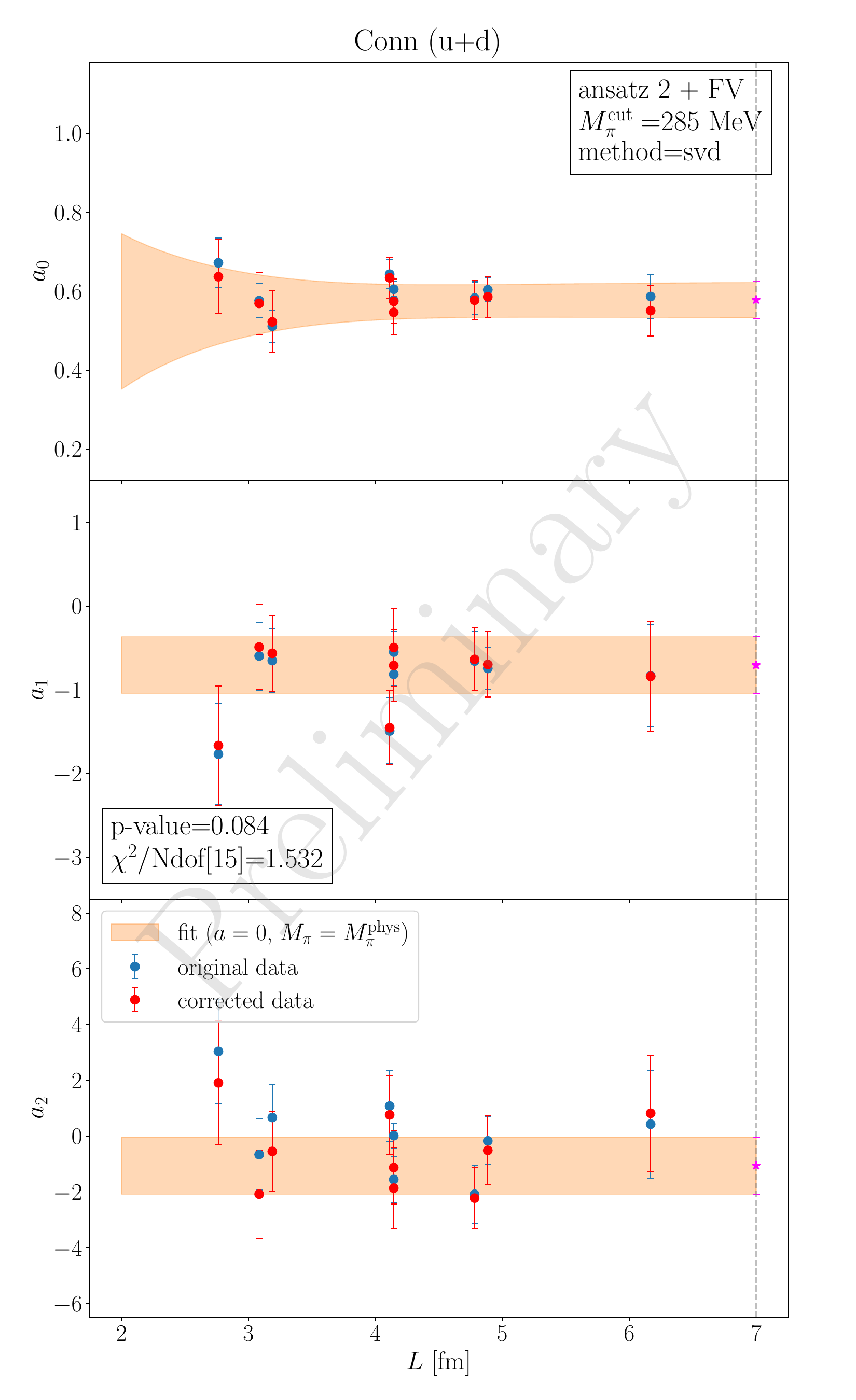}
}
\caption{Example of chiral-continumm extrapolation for the $u+d$ case with ansatz 2 and finite-volume effect with a cut $M^{\rm cut}_{\pi}=285\,\text{MeV}$, shown
for all the three coefficients (rows)
 as a function of $M_{\pi}^2$ (left), $a^2$ (centre) and the spatial lattice size $L$ (right). The blue points are the original data and the red points and band correspond to the corrected version for the continuum parameters as specified in the legend.}
\label{fig:chiral-continuum}
\end{figure}

In this section we highlight the final steps of the analysis on the connected data, namely the chiral-continuum extrapolation of the coefficients of the $z$-expansion
and the model average, and compare with preliminary results
on the full dataset of two of our most chiral ensembles.

We consider three ans\"atze:
\begin{enumerate}
 \item linear in $M_{\pi}^2$ and $a^2$ for each coefficient $a_i$;
 \item same as ansatz 1 with the addition of an $M_{\pi}^3$ term and a log term for the axial charge $a_0$;
 \item same as ansatz 2 with the addition of $M_{\pi}^3$ terms for $a_1$ and $a_2$. 
\end{enumerate}
To account for finite-volume effects, we also consider all the previous ans\"atze with the correction term
\begin{align}
 \frac{M_\pi^2}{\sqrt{M_\pi L}}e^{-M_\pi L}  
\end{align}
for $a_0$. We show an example of the chiral-continuum extrapolation of the connected data with ansatz 2 and finite-volume effects  in~\figref{fig:chiral-continuum}, as a function of $M_{\pi}^2$,
$a^2$ and the spatial lattice size $L$. The plot shows that the behaviour is quite flat for all the three variables, suggesting that the simple ansatz 1 would be enough to describe the data. In addition, finite-volume effects appear to be negligible.

\begin{figure}[t]
\centering
\includegraphics[scale=0.3]{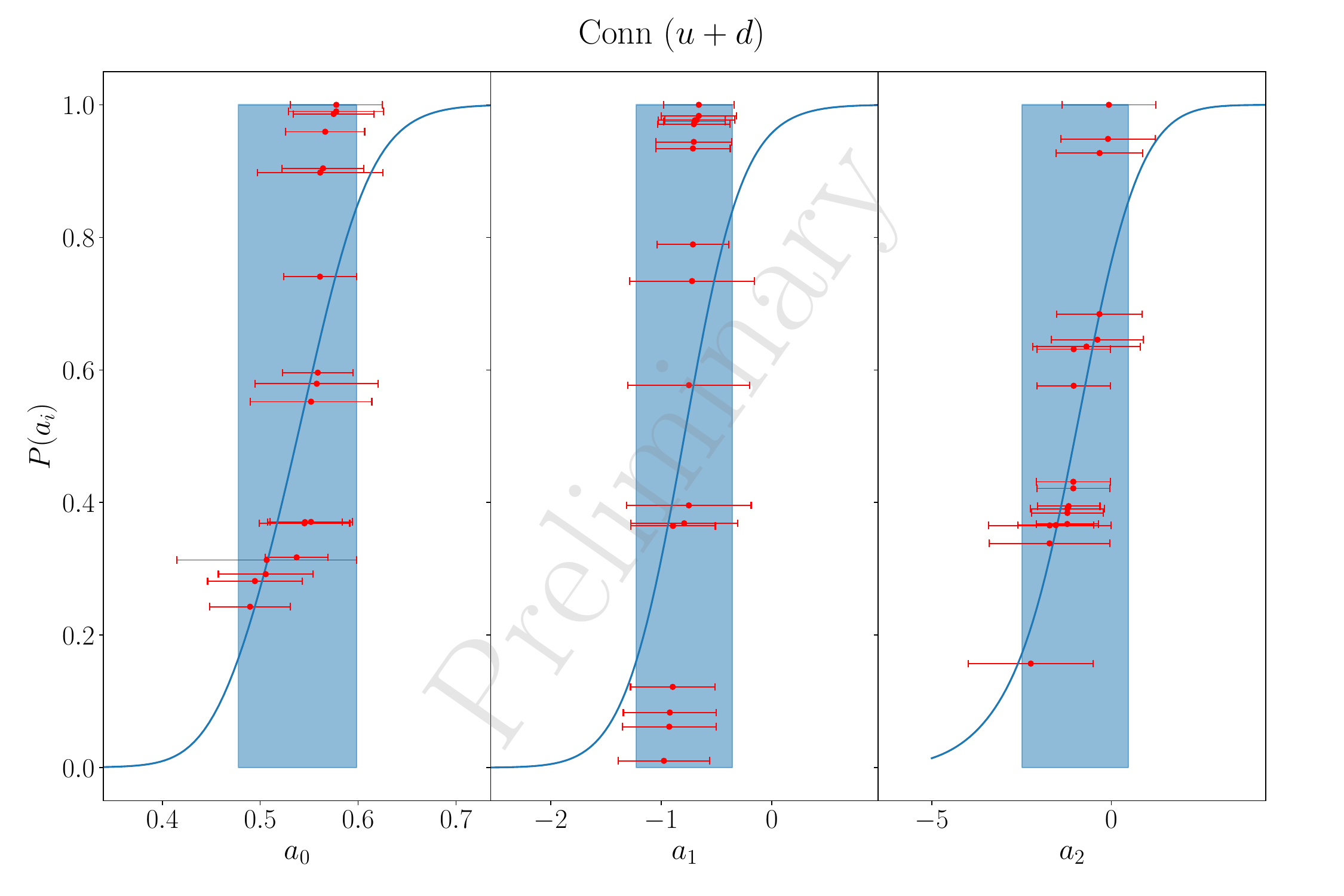}
\caption{Model average through AIC for the $u+d$ case. The red points correspond to the results of the fits entering the model average, and the blue line is
the cumulative distribution in~\eqrefeq{eq:cumulative}, with the vertical bands indicating the final results obtained from the 16th and 84th percentiles.} 
\label{fig:AIC}
\end{figure}

We perform multiple fits with cuts in the pion mass, $M^{\rm cut}_{\pi}[\text{MeV}]=\{300, 285, 265\}$, and cuts in the coarsest lattice spacing, while preserving the correlations
among the three coefficients on each ensemble. We then obtain the final result 
through model average exploiting the version of the Akaike Information Criterion proposed in~\cite{Borsanyi:2020mff}, i.e. assigning to each $k$-th fit the weight
\begin{align}
 w^{\rm AIC}_k \propto 
 e^{-\frac{1}{2}(\chi^2_k +2n_{{\rm par}, k} -n_{{\rm data}, k} )}  \, ,
\end{align}
with $n_{{\rm par}, k}$ being the number of parameters and $n_{{\rm data}, k}$ the number of data points entering the fit,
and obtaining the final results exploiting the 16th, 50th and 84th percentiles of the cumulative distributions
\begin{align}
\label{eq:cumulative}
 P(a_i) = \int_{-\infty}^{a_i} \dd a_i' \sum_{k} w^{\rm AIC}_k
 \mathcal{N}(a_i'; \langle a_i^{(k)} \rangle, \sigma_{a_i^{(k)}})
\end{align}
obtained from a weighted sum of normal distributions centered on $\langle a_i^{(k)} \rangle$ and variance $\sigma^2_{a_i^{(k)}}$
for each $z$-expansion coefficient $a_i$ and fit $k$. The procedure is shown in~\figref{fig:AIC}. The correlations are taken into account repeating the same procedure
for the cumulative distributions $P(a_i a_j)$ and extracting them from the standard relations between the variances $\text{var}[a_ia_j]$, $\text{var}[a_i]$ and $\text{var}[a_j]$.

\begin{figure}
\centering
\includegraphics[scale=0.33]{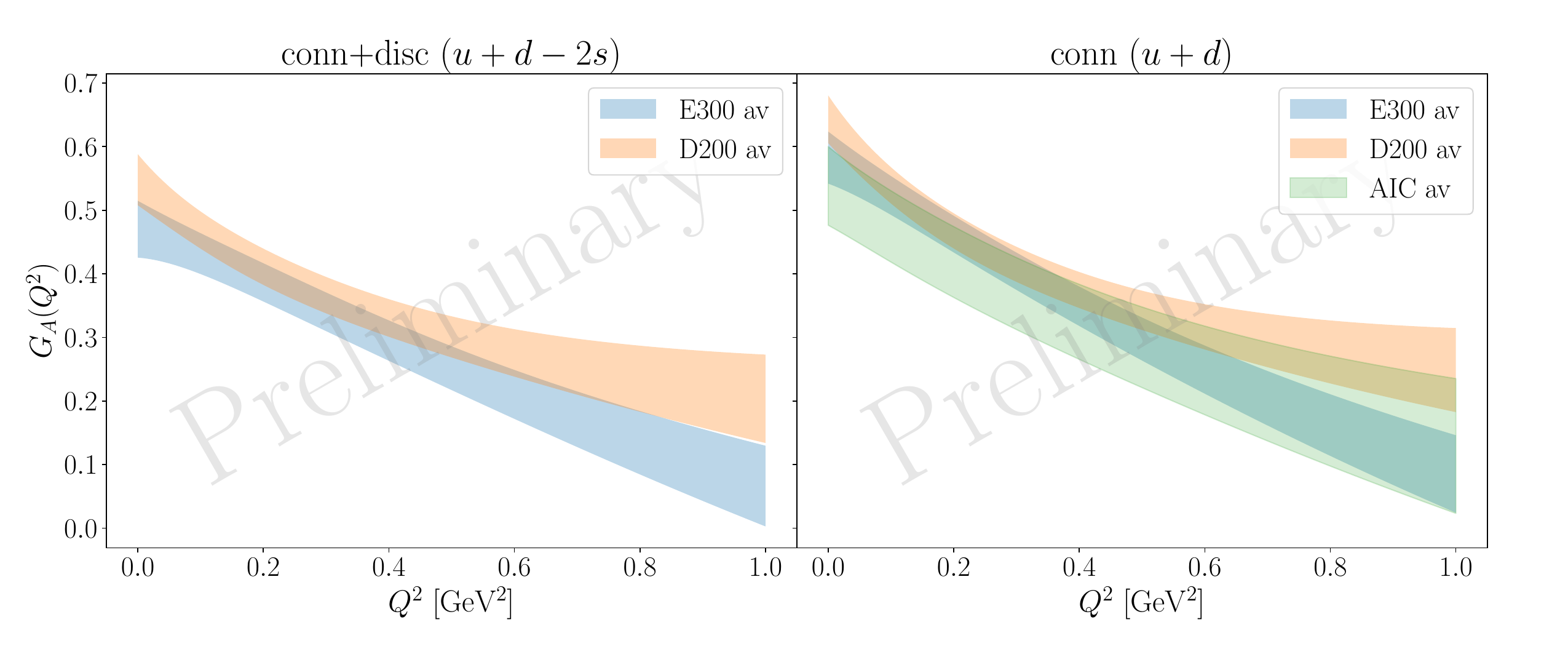}
\caption{Final results on the isoscalar axial form factor $G_A(Q^2)$ on the ensembles E300 and D200 after window average for both connected (right) and full case (left), compared with
the final AIC average for the connected case only in green.}
\label{fig:final}
\end{figure}

We illustrate the preliminary results in~\figref{fig:final}, where we compare the connected $u+d$ contribution
after model average to two of the most chiral ensembles E300 and D200 (right); the full $u+d-2s$
from factor is displayed on the left for these two ensembles.
We can see that for the connected case the two ensembles already seem to provide a good description of the physical case.
The disconnected pieces contribute mainly at low $Q^2$ (cf.~\figref{fig:zfit-comparison}) yielding a shift of the form factor mainly in that region.
While this preliminary evaluation is obtained only on two ensembles, i.e without accounting for lattice artifacts, these provide a value of the axial charge compatible 
with $g_A^{u+d-2s}=0.46(5)$ obtained using the Cloudy Bag model~\cite{Bass:2009ed}, as well as the most recent result from the ETM collaboration
$g_A^{u+d-2s}=0.490(20)$~\cite{Alexandrou:2024ozj}.

\section{Outlook and conclusions}

In this proceedings contribution we have outlined our analysis strategy for the non-singlet isoscalar axial form factor, reporting some preliminary results for both the connected and the full case on a few ensembles. In particular, we exploit the summation method combined with a direct $z$-expansion to order $n=2$, comparing various techniques to regulate the large covariance matrix, namely off-diagonal damping and svd cuts. The $z$-expansion coefficients are obtained from each ensemble through a window average of the minimum source-sink separations - which have been kept in physical units across all ensembles in order to reduce the human bias - and then extrapolated to the chiral-continuum limit with different
ans\"atze and cuts both in pion mass and lattice spacing. The final result is then obtained through a model average.

To complete the analysis, several steps have to be taken. First of all, we will include the disconnected contributions on all ensembles
to extend the analysis to the full $u+d-2s$ case.
We plan to explore more fit ans\"atze (e.g. dipole) to cross-check the quality of our data and the performance of our analysis.
Once complete, this study will provide a first physical result for the isoscalar octect of the nucleon axial form factor in a large $Q^2$ range accessible by experiments. Furthermore,
it will provide a first step into the flavour decomposition of the form factor, for which we require a similar analysis for the singlet $u+d+s$ contribution.

\section*{Acknowledgments}

 This work was
  supported in part by the European Research Council (ERC) under the
  European Union’s Horizon 2020 research and innovation program
  through Grant Agreement No.\ 771971-SIMDAMA and by the Deutsche
  Forschungsgemeinschaft (DFG) through the Collaborative Research
  Center  1660 ``Hadrons and Nuclei as Discovery Tools'',
  under grant HI~2048/1-3 (Project No.\ 399400745) and in the Cluster
  of Excellence “Precision Physics, Fundamental Interactions and
  Structure of Matter” (PRISMA+ EXC 2118/1) funded by the DFG within
  the German Excellence strategy (Project ID 39083149).
  Calculations for this project were partly performed on the HPC
  clusters ``Clover'' and ``HIMster2'' at the Helmholtz Institute Mainz,
  and ``Mogon 2'' at Johannes Gutenberg-Universit\"at Mainz.
  The authors gratefully acknowledge the Gauss Centre for Supercomputing e.V. (www.gauss-centre.eu) 
  for funding this project by providing computing time on the GCS Supercomputer systems JUQUEEN and JUWELS at J\"ulich Supercomputing Centre (JSC) 
  via grants HMZ21, HMZ23 and HMZ36 (the latter through the John von Neumann Institute for Computing (NIC)), as well as on the GCS Supercomputer HAZELHEN at
   H\"ochstleistungsrechenzentrum Stuttgart (www.hlrs.de) under project GCS-HQCD.
  
Our programs use the QDP++ library~\cite{Edwards:2004sx} and deflated SAP+GCR
solver from the openQCD package~\cite{Luscher:2012av}, while the contractions
have been explicitly checked using~\cite{Djukanovic:2016spv}. We are grateful to
our colleagues in the CLS initiative for sharing the gauge field configurations
on which this work is based.

\bibliographystyle{JHEP}

\providecommand{\href}[2]{#2}\begingroup\raggedright\endgroup

\end{document}